\documentclass[10pt]{amsart}
\usepackage{amsmath}
\usepackage{amssymb}
\usepackage{yfonts}

\usepackage{tikz,pgfplots}

\usetikzlibrary{calc, patterns,arrows, shapes.geometric}

\newtheorem{theorem}{Theorem}[section]
\newtheorem{proposition}[theorem]{Proposition}

\begin{document}

\title{Constrained ballistics and geometrical optics}
\author{Marcelo Epstein}

\address{Department of Mechanical and Manufacturing Engineering, University of Calgary, Canada}
\email{mepstein@ucalgary.ca}


\date{}
\maketitle

\begin{abstract}
The problem of constant-speed ballistics is studied under the umbrella of non-linear non-holonomic constrained systems. The Newtonian approach is shown to be equivalent to the use of Chetaev's rule to incorporate the constraint within the initially unconstrained formulation. Although the resulting equations are not, in principle, obtained from a variational statement, it is shown that the trajectories coincide with those of geometrical optics in a medium with a suitably chosen refractive index, as prescribed by Fermat's principle of least time. This fact gives rise to an intriguing mechano-optical analogy. The trajectories are further studied and discussed.
\end{abstract}

\section{Introduction}

A missile is launched in a gravitational field with an initial velocity ${\bf v}_0$. What will its trajectory be if its speed is forced to remain constant? More generally, assume that we are given an a-priori specification of the Lagrangian $L$ of an unconstrained system. If we work in a coordinate representation, let the generalized coordinates be denoted by $q^i\;(i=1,...,N)$. Then, we assume the Lagrangian function to be given by
\begin{equation} \label{speed5}
L=L(q^i,{\dot q}^i,t),
\end{equation}
where $t$ is time and a superimposed dot denotes the total time-derivative. Assume now that the following information has been received: The previously unconstrained system is to be subjected to a constraint of the form
\begin{equation} \label{speed6}
G(q^i, {\dot q}^i,t)=0,
\end{equation}
where $G$ is a smooth function of its arguments. The question we address is: What modifications need to be effected to the Lagrangian formulations so as to represent the dynamics of the new constrained system? Notice that the fact that the evolution of the original unconstrained system was represented by the stationarity of a certain integral may turn out to be quite irrelevant for the dynamics of the new system.

In particular, for our ballistic problem, the Lagrangian is
\begin{equation} \label{speed6a}
L= \frac{1}{2}m ({\dot x}_i {\dot x}_i) - U(x_1,x_2,x_3)
\end{equation}
where $m$ is the mass of the particle, $U$ is the gravitational potential and $x_1,x_2,x_3$ are Cartesian coordinates in an inertial frame. The summation convention for repeated indices is in force. The constant-speed constraint is
\begin{equation} \label{speed7}
{\dot x}_i {\dot x}_i - v_0^2=0.
\end{equation}
This is a nonlinear nonholonomic constraint.

There is no standard way to account for constraints in Analytical Mechanics. In the holonomic case one often resorts to the following additional information: The constraint is workless. But it is easy to generate examples, even of cases involving no friction, in which a properly described holonomic constraint is not workless. A good example is provided by the strategy of holding an inverted broom stably in its vertical position on your fingertip by just moving the finger along a horizontal line. Assuming no rotational friction, it is not difficult to show that the system can be kept in a vertical position of stable equilibrium by imposing the holonomic constraint $x-C\theta=0$, where $x$ and $\theta$ represent, respectively, the horizontal displacement of the fingertip and the angular deviation from the vertical, and where the constant $C$ is large enough. By resorting to a purely Newtonian formulation, it can be shown that this constraint results in a constant active removal of energy from the system by means of the forced small-amplitude oscillations around the stable equilibrium position. But how is the Newtonian solution replicated in the Lagrangian context? It turns out that there is no clear-cut canonical way to incorporate the constraint into the Lagrangian framework. In other words, the constraint equation alone is not sufficient to pin down the Lagrange equations of motion of the constrained system, but more information needs to be provided.

In the case of nonholonomic constraints, this lack of determination is further exacerbated by the fact that even the notion of `workless' constraint is not applicable in general. The closest one gets to this notion is the content of the so-called Chetaev rule. On the other extreme, there exists a point of view that advocates the use of `vakonomics' to preserve the variational character of the formulation. Both of these prescriptions apply to systems that, were not for the presence of the constraint, would abide by a principle of stationarity of a given Lagrangian functional. We will study the implications of these two extreme formulations to the case of constrained ballistics. A surprising physical by-product of this analysis is that the Chetaev trajectories (which are not associated, in principle, with the stationarity of a functional) turn out to be identical to the trajectories of a ray of light (governed by Fermat's principle) in a refractive medium whose index of refraction is determined by the gravitational potential of the projectile motion. This is a remarkable mechano-optical analogy worthy of further study.

\section{Newtonian analysis}

If we assume the speed of the projectile to be controlled by an active mechanism that does not change the mass of the system, a differentiation of the constraint (\ref{speed7}) yields
\begin{equation} \label{7a}
{\dot x}_i {\ddot x}_i = 0.
\end{equation}
This equation can be interpreted as follows: The total force acting on the missile (being, according to Newton's laws, proportional to its acceleration) is constantly perpendicular to the trajectory. In other words, as expected, the tangential acceleration vanishes. It follows, therefore, that the mechanism of speed control must apply a force of varying magnitude in the direction of the trajectory (by, say, a computer-controlled jet engine). Denoting the scalar measure of this force by $F$, the Newtonian equations of motion are
\begin{equation} \label{7b}
m{\ddot x}_i+\frac{\partial U}{\partial x_i} - F {\dot x}_i = 0.
\end{equation}

Multiplying Equations (\ref{7b})  by ${\dot x}_i$,
it follows that, by virtue of (\ref{speed7}),
\begin{equation} \label{7d}
F=\frac{1}{v_0^2}\; \frac{\partial U}{\partial x_i}{\dot x}_i = \frac{1}{{\dot x}_k{\dot x}_k}\; \frac{\partial U}{\partial x_i}{\dot x}_i.
\end{equation}
Having thus determined the force field consistent with the constraint, the Newtonian formulation is complete and unequivocal. In particular, it can be verified by direct substitution that ${\dot x}_k{\dot x}_k$ is a constant of the motion.

\section{Chetaev's rule}

Chetaev's rule postulates that the equations of motion of the constrained system defined by Equations (\ref{speed5}) and (\ref{speed6}) are given by
\begin{equation} \label{speed8}
\frac{\partial L}{\partial q^i}-\frac{d}{dt}\left(\frac{\partial L}{\partial {\dot q}^i}\right) - \lambda \frac{\partial G}{\partial{\dot q}^i} = 0.
\end{equation}
These $N$ equations, when supplemented by the constraint equation itself, furnish a set of $N+1$ ODEs for the evolution of the $N$ functions $q^i(t)$ and the additional function $\lambda(t)$, known as the Lagrange multiplier associated with the constraint.

For the particular case of Equations (\ref{speed6a}) and (\ref{speed7}), we obtain
\begin{equation} \label{speed9}
m{\ddot x}_i+ \frac{\partial U}{\partial x_i} + 2\lambda {\dot x}_i=0.
\end{equation}
Comparing with Equation (\ref{7b}), we conclude that, with the identification $F=-2\lambda$, Chetaev's rule leads to the same result as the Newtonian formulation based on the mass constancy. Notice that the nonholonomicity of the constraint destroys, in general, the variational character of the formulation, in the sense that it does not correspond to the stationarity of a functional.

\section{A remarkable mechano-optical analogy}

The propagation of light rays in an isotropic refractive medium is governed by Fermat's principle of least time. It states that the integral
\begin{equation} \label{opt1}
T = \int\limits_A^B \frac{n}{c}\; ds
\end{equation}
representing the time of travel of a light ray between two fixed points, $A$ and $B$, is stationary with respect to all paths passing through these points. In this equation, $c$ represents the speed of light in vacuo, $n=n(x_1,x_2,x_3)$ is the (non-dimensional) index of refraction and $ds$ is the length element along the trajectory. Indicating with primes the derivatives with respect to an arbitrary curve parameter $\tau$, we write
\begin{equation} \label{opt2}
T = \int\limits_A^B \frac{n}{c}\;\sqrt{x_i' x_i'}\; d\tau,
\end{equation}
whose Euler-Lagrange equations are given by the system
\begin{equation} \label{opt3}
\frac{1}{c}\;\frac{\partial n}{\partial x_i}\;\sqrt{x_k' x_k'} - \frac{d}{d\tau}\left(\frac{n}{c}\; \frac{x_i'}{\sqrt{x_k' x_k'}}\right) = 0.
\end{equation}

Since the integrand in the functional (\ref{opt2}) is homogeneous of degree 1, we know \cite{courant} that the formulation is independent on the choice of parameter. Consequently, we are in our right to choose the parameter $\tau=s/v_0$, which is proportional to the arc-length parameter $s$. In this case, since $x_i' x_i'=v_0^2$, the Euler Lagrange equations can be written as
\begin{equation} \label{opt4}
\frac{\partial n}{\partial x_i} -\frac{1}{v_0^2}\; \frac{d}{d\tau}\left(n\; x_i'\right) = 0,
\end{equation}
or, equivalently,
\begin{equation} \label{opt5}
x_i''+\frac{\partial \ln{ n}}{\partial x_k}\;{x_k' x_i'} - \frac{\partial \ln{ n}}{\partial x_i'} = 0.
\end{equation}
Setting the refractive index to
\begin{equation} \label{opt6}
n=e^{-\frac{U}{m v_0^2}},
\end{equation}
Equation (\ref{opt6}) becomes identical to (\ref{7b}) or (\ref{speed9}). We have thus proven the following proposition.

\begin{proposition} {\bf The mechano-optical analogy}: The ballistic trajectories under a constant-speed constraint are identical to the trajectories of light rays in a refractive medium whose refractive index is obtained from a given gravitational potential according to Equation (\ref{opt6}).
\end{proposition}

\section{Explicit ballistic trajectories}

For a constant gravitational field $-g$ acting in the vertical direction $y$, assuming the motion to take place in the $x,y$ plane, the equations of motion become
\begin{equation} \label{traj1}
{\ddot x} - \frac{g}{v_0^2}\;{\dot x}{\dot y} = 0,
\end{equation}
and
\begin{equation} \label{traj2}
{\ddot y} - \frac{g}{v_0^2}\;{\dot y}^2 + g = 0.
\end{equation}
Introducing the characteristic time $T_c=v_0/g$ and the characteristic length $L=v_0 T_c$, we define non-dimensional time and space coordinate as
\begin{equation} \label{traj3}
\zeta=\frac{t}{T_c},\;\;\;\;\;\;\xi=\frac{x}{L},\;\;\;\;\;\; \eta = \frac{y}{L},
\end{equation}
and rewrite the equations of motion as
\begin{equation} \label{traj4}
\xi''-\xi' \eta' = 0,
\end{equation}
and
\begin{equation} \label{traj5}
\eta'' - \eta'^2 -1=0,
\end{equation}
where primes now denote $\zeta$-derivatives.

The general solution of this system can be expressed as
\begin{equation} \label{traj6}
\xi(\zeta)=2\arctan\left(\frac{e^\zeta}{\sqrt{B}}\right) +C,
\end{equation}
and
\begin{equation} \label{traj7}
\eta(\zeta)= \zeta - \ln \left(e^{2\zeta}+B\right) +A.
\end{equation}
The constant $B$ is always non-negative and is related to the initial slope by
\begin{equation} \label{traj8}
\frac{B-1}{2\sqrt{B}}=\left(\frac{d\eta}{d\xi}\right)_{\zeta=0}.
\end{equation}
Note tha $B=0$ corresponds to direct downward fall, while $B\to \infty$ indicates direct vertical ascent. Horizontal firing corresponds to $B=1$. Without loss of generality, we may assume that $A=C=0$ and determine thereby the initial location.

All trajectories have a finite range, unlike the unconstrained case where (provided there is enough free depth) any distance is horizontally reachable. To calculate this range we evaluate
\begin{equation} \label{traj9}
\lim_{\zeta \to \infty} \xi(\zeta) = \pi.
\end{equation}
The whole trajectory is, therefore, contained in the interval ($0,\pi$). As $\zeta \to \infty$, the trajectory attains a vertical slope and proceeds to infinite depth. In terms of the original dimensional variables, the maximum range $R$ is
\begin{equation} \label{traj10}
R=\frac{\pi v_0^2}{g}.
\end{equation}
as a curiosity, we recall the classical Torricelli height of an unconstrained vertically launched projectile, namely $h=v_0^2/2g$, which shows that the horizontal range of the constrained projectile is only about 6 times larger. Finally, we note that, eliminating the time-variable $\zeta$, the shape of the trajectory is given by the function
\begin{equation} \label{traj11}
\eta=\ln\left(\frac{\sin \xi}{2\sqrt{B}}\right),
\end{equation}
which is concave and whose maximum occurs at $\xi=\pi/2$. A projectile launched with a positive initial slope $s_0$ will attain a maximal height of
\begin{equation}
h_{max}=\frac{v_0^2}{g}\;\ln\left(\frac{1+s_0^2 - s_0\sqrt{s_0^2+1}}{s_0+\sqrt{s_0^2+1}}\right).
\end{equation}

\end{document}